\documentclass[a4paper, amsfonts, amssymb, amsmath, reprint, showkeys, nofootinbib,twocolumn,dvipsnames,superscriptaddress]{revtex4-2} 

\usepackage{graphicx}
\usepackage{dcolumn}
\usepackage{bm}
\usepackage{mathrsfs}
\usepackage{amsmath,amsthm,amssymb,amsfonts}
\usepackage{dsfont}
\usepackage{xcolor}
\usepackage{tikz}
\usepackage{babel}
\usetikzlibrary{shapes.geometric} 

\usepackage{mathabx}
\usepackage{braket}
\usepackage{comment}
\usepackage{enumitem}
\usepackage[pdftex, pdftitle={Article}, pdfauthor={Author}]{hyperref} 
\usepackage[utf8]{inputenc}
\usepackage[colorinlistoftodos, color=green!40, prependcaption]{todonotes}
\usepackage{multirow}
\usepackage{ulem}
\usepackage[left=13mm,right=13mm,top=25mm,bottom=25mm,columnsep=15pt]{geometry}

\setlist[enumerate]{itemsep=0mm} 
\usetikzlibrary{quantikz2} 

\definecolor{darkgreen}{rgb}{0.0,0.5,0.0}

\usepackage{tabularx}

\newcounter{protocol}
\newenvironment{protocol}[1]
  {\par\addvspace{\topsep}
   \noindent
   \tabularx{\linewidth}{@{} X @{}}
    \hline
    \refstepcounter{protocol}\textbf{Protocol \theprotocol} #1 \\
    \hline}
  { \\
    \hline
   \endtabularx
   \par\addvspace{\topsep}}

\newcommand{\sbline}{\\[.5\normalbaselineskip]}

\usepackage{tikzpeople} 
\usetikzlibrary{shapes, arrows.meta, positioning}
\usetikzlibrary{positioning}

\begin{document}

\title{Experimental Quantum Electronic Voting}

\author{Nicolas Laurent-Puig}
    \email[Correspondence email address:]{nicolas.laurent-puig@lip6.fr}
    \affiliation{Sorbonne Université, CNRS, LIP6, 4 Place Jussieu, Paris F-75005, France}
\author{Matilde Baroni}
    \affiliation{Sorbonne Université, CNRS, LIP6, 4 Place Jussieu, Paris F-75005, France}
\author{Federico Centrone}
    \affiliation{ICFO-Institut de Ciencies Fotoniques, The Barcelona Institute of Science and Technology,
Av. Carl Friedrich Gauss 3, 08860 Castelldefels (Barcelona), Spain.}
\author{Eleni Diamanti}
    \affiliation{Sorbonne Université, CNRS, LIP6, 4 Place Jussieu, Paris F-75005, France}

\date{\today} 
\begin{abstract}
Quantum information protocols offer significant advantages in properties such as security, anonymity, and privacy for communication and computing tasks. An application where guaranteeing the highest possible security and privacy is critical for democratic societies is electronic voting. As computational power continues to evolve, classical voting schemes may become increasingly vulnerable to information leakage. In this work, we present the experimental demonstration of an information-theoretically secure and efficient electronic voting protocol that, crucially, does not rely on election authorities, leveraging the unique properties of quantum states. Our experiment is based on a high-performance source of Greenberger–Horne–Zeilinger (GHZ) states and realizes a proof-of-principle implementation of the protocol in two scenarios: a configuration with four voters and two candidates employing privacy enhancement techniques and an election scenario supporting up to eight voters and sixteen candidates. The latter is particularly well-suited for secure board-level elections within organizations or small-scale governmental contexts.

\end{abstract}

\maketitle

\textbf{Introduction.}\label{sec:Introduction}
Electronic voting offers, in principle, a powerful alternative to traditional election systems, enabling secure and verifiable large-scale elections even in the presence of dishonest authorities. Yet, defining rigorous security properties for such protocols remains a substantial challenge~\cite{ChevallierMames2010}. Classical e-voting protocols often require replacing physically grounded assumptions—such as ballot boxes or private booths—with cryptographic ones, which can be especially problematic in remote settings where trust is inherently limited. Despite these challenges, real-world electronic voting systems have been deployed. One notable example is the system used in France during the June 2022 legislative elections for overseas residents, involving 1.6 million eligible voters. Derived from an academic protocol~\cite{Cortier2019}, this platform was used to elect 11 deputies, collecting over 524,000 electronic ballots, making it the largest e-voting deployment to date in terms of ballots cast. However, subsequent reverse engineering by French researchers uncovered critical vulnerabilities in the system~\cite{debant:hal-03875463}, raising serious concerns about its long-term security. Although information-theoretic e-voting protocols have been proposed~\cite{10.1007/3-540-68339-9_7}, they often rely on strong assumptions such as trusted authorities or simultaneous broadcast channels—both of which are impractical in many real-world scenarios. Post-quantum cryptography may offer a more robust alternative~\cite{10.1007/978-3-662-63958-0_25}, although long-term guarantees remain unsettled.

Over the past decade, quantum electronic voting has emerged as a potential candidate for addressing these foundational limitations~\cite{PhysRevA.84.022331,PhysRevA.94.022333,1304.0555}. However, several of the earlier proposals suffer from serious flaws~\cite{d1c2789bf9ca4fa2a3c886e7d73e44c1}. More recently, an information-theoretically secure quantum e-voting protocol introduced by some of the authors~\cite{Centrone_2022} showed that it was possible to avoid reliance on trusted authorities while being within experimental reach with current quantum information technologies. The scheme leverages multipartite entanglement, where each of the $N$ agent receives a qubit from an $N$-qubit GHZ state, and can perform local measurements. The protocol incorporates quantum verification and classical anonymous transmission subroutines, ensuring both privacy and correctness.

In this work, we enhance and extend the previously proposed quantum e-voting protocol and provide an experimental implementation that demonstrates the security and privacy guarantees it offers in realistic scenarios. More specifically, on the theoretical side, we rigorously introduce privacy enhancement techniques and extend the protocol to support multiple candidates and voting pools, thereby greatly improving its efficiency and practical relevance. Technically, this relies on the efficient composition of repeated executions of the original protocol. On the experimental side, we leverage a compact, high-fidelity GHZ source operating at telecom wavelengths~\cite{martins2024realizingcompacthighfidelitytelecomwavelength}, making the implementation compatible with existing communication infrastructure. We demonstrate in practice two distinct election scenarios: a two-candidate election with four voters, and a multi-candidate election involving two independent pools of four voters—hence, highlighting the scalability and practical viability of the protocol. It is important to note that although our setup does not include quantum memories—an essential component for implementing the full protocol—such devices are continuously gaining in performance and maturity~\cite{ScienceAdvances2025}, keeping our demonstration aligned with the capabilities of near-term quantum technologies. Our results demonstrate, for the first time to our knowledge, a quantum electronic voting protocol operating under realistic conditions, and take a step toward the practical deployment of quantum-secure democratic election procedures.

\textbf{Theoretical protocol.}\label{sec:{Theoretical protocol}}
We present in the following the full quantum e-voting protocol (Protocol 1) as well as the \texttt{Voting} subroutine (Protocol 2). Key aspects of the protocol include the implementation of privacy enhancement and the support of multiple candidates and multiple pools of voters in an efficient way. Both are based on the composition of many repetitions of the core protocol. \texttt{Verification}, which is the same as in the original protocol, is provided in End Matter~\ref{sec:Verification}, and the various classical subroutines are provided in Supplementary Materials~\cite{supp}.
The role of \texttt{Verification} is crucial: in the protocol, a central source of $N$‑qubit GHZ states, $\ket{\mathrm{GHZ}} = \frac{1}{\sqrt{2}}\Bigl(\ket{0}^{\otimes N} + \ket{1}^{\otimes N}\Bigr)$, distributes a qubit to each of the $N$ agents, who must be able to receive one photon, temporarily store the photon, apply a local unitary transformation, and perform a measurement. Thanks to this subroutine, agents do not need to trust the source, as it ensures that the distributed state has a fidelity above a fixed threshold. 

\begin{protocol}{Quantum e-voting}
\vspace{1pt}
\textit{Goal.} $N$ agents anonymously express their preference between $C$ candidates.
\sbline
\textit{Inputs.} For all $n \in [N]$, voter holds an input~$s_n \in [C]$.
$\Pi$ rounds of privacy enhancement (PE).
Security parameter $S$ for the classical subroutines.
$M$ number of random coins.
$\delta$ threshold for the verification.
\sbline
\textit{Outputs.} The tally of the election \textbf{T}.
\sbline
\textit{Resources.}  $N$-partite GHZ source, single-qubit operations, classical and quantum communication, random numbers.
\sbline
\textit{The protocol:}

\begin{enumerate}
  \item \textbf{Secret orderings.}
    The agents perform \texttt{UniqueIndex}, and everyone obtains a unique secret random index $\omega_n \in [N]$.
  \item \textbf{Cast of the votes.}
\par For $k=1$ to $\lceil \log_2(C)\rceil$ [Digit round]:
\par  For $p =1$ to $\Pi$ [PE round]:
  \par For $n =1$ to $N$:
    \begin{enumerate}
    \item all agents $j \in [N]$ set $\texttt{r}_j=\texttt{t}_j$=0, and receive a qubit of the GHZ state from the source;
    \item the voting agent $a= \omega_n \in[N]$ samples a random variable $ r\leftarrow \textsf{Bernoulli}(2^{-M})$, and anonymously announces the result using \texttt{RandomBit};
    \begin{enumerate}
        \item if $r=0$, it is a \texttt{Verification} sub-subround:
        \begin{enumerate}
            \item the voting agent $a= \omega_n$ performs \texttt{RandomAgent} to randomly select a verifier $v$, $\texttt{t}_v=\texttt{t}_v+1$;
            \item  the agent $v$ performs \texttt{Verification}, and if it outputs reject $\texttt{r}_v=\texttt{r}_v+1$;
            \item go back to 2 (b);
        \end{enumerate}
        \item if $r=1$, it is a \texttt{Voting} sub-subround :
        \begin{enumerate}
            \item for all $j \in [N]$ compute $\delta_j= \frac{\texttt{r}_j}{\texttt{t}_j}$;
            \item if for any $j$ $\delta_j > \delta$, go back to 2 (a), otherwise perform \texttt{Voting}.
        \end{enumerate}
    \end{enumerate}
  \end{enumerate}
   \item \textbf{Create the tally.} All agents agree on the bulletin board \textbf{B}.
   Compute the XOR over each row, obtaining a $N \times \lceil\log_2(C)\rceil$ vertical vector; from this it is straightforward to compute the tally. Return the tally \textbf{T}.

\end{enumerate}
\end{protocol}

\begin{protocol}{\texttt{Voting}}
\vspace{1pt}
\textit{Goal.} Express anonymously the preference of the voting agent $a=\omega_n \in[N]$.
\sbline
\textit{Inputs.} The subround counter $(p,k,n)$, that identifies the voting agent $a=\omega_n$; the $k$-th bit of the input $s_a\in[C]$ represented in binary units, $s_a^k \in \{0,1\}$.
\sbline
\textit{Outputs.} The result of a subround, which is a $N$-dimensional vector that will be part of the bulletin board $B$, more precisely the $n$-th row of the bulletin of the $(p,k)$-subround.
\sbline
\textit{The protocol:}
\begin{enumerate}
 \item \textbf{Measurements.} Each voter $j\in[N]$ measures their state on the Hadamard basis, and records the outcome $d_j^{p,k,n} \in \{0,1\}$.  
 \item If $p < \Pi$: \textbf{Encode randomness.} The voting agent $a$ samples a random bit $r_a$, and sets $d_a^{p,k,n}= d_a^{p,k,n} \oplus r_a$.
 \item Else if $p=\Pi$: \textbf{Encode the vote.} The voting agent $a$ fixes the parity of their result to be exactly $s_a^k$, by setting  $d_a^{\Pi,k,n}= s_a^k \bigoplus_{p=1}^{\Pi-1} d_a^{p,k,n}$.
    \item \textbf{Broadcast the result.} Every agent publicly broadcasts their result; the vector collecting all the results $\{d_j^{(p,k,n)}\}_{j\in[N]}$is the $n$-th row of the $(p,k)$-subround.
\item Allow aborting:
\begin{enumerate}
        \item for each party $n \in [N]$, if their intention is correctly encoded in the final vector of outcomes $c_n=0$, otherwise $c_n=1$;
        \item perform a \texttt{LogicalOR} with the $\{c_i\}_i$:
        \begin{enumerate}
            \item if it outputs $0$, do nothing;
            \item else if it outputs $1$, go back to 2 (a).
        \end{enumerate}
        \end{enumerate}
\end{enumerate}
\end{protocol}

We now provide a detailed description of Protocol 1, which consists of three phases. In the first phase,
a secret index \(\omega_n\) is assigned to each agent via the \texttt{UniqueIndex} classical subroutine, which ensures that each agent only knows when they are scheduled to vote, while the global scheduling is kept secret.

The second phase involves multiple repetitions of a core procedure, which itself consists of several rounds of verification followed by a vote. In the end, all parties agree on a bulletin board, which is the collection of many sub-bulletin boards, as represented in Fig.~\ref{fig:bulletin}. 

The third and final phase is a simple post-processing of these results to obtain from the bulletin the tally and the final voting result.

Let us focus more on the second phase of the protocol.
This is composed of three nested ``for'' loops: a label $k \in [\lceil \log_2(C)\rceil ]$ loops over the digits of the admissible votes, a label $p \in [\Pi]$ keeps track of the privacy enhancement rounds, and a label $n\in [N]$ of all the agents. We first focus on a loop for a fixed value of $k$ and $p$, which is called a subround.
In each subround, the voting agent randomly chooses to execute either \texttt{Verification} or \texttt{Voting} subroutine by throwing \(M\) random coins, and voting if and only if all the coins are heads. This random variable follows a Bernoulli distribution with probability of success \(2^{-M}\); \(M\) is a security parameter that can be tuned to ensure that \texttt{Verification} is performed with high frequency to check the integrity of the shared GHZ state.

\noindent\texttt{Verification.} One agent is chosen uniformly at random to coordinate the verification. This coordinator distributes random rotation angles \(\theta_n\), and increases her verification count $t_i$ by 1. All agents measure their qubit on the Hadamard basis rotated by \(\theta_n\). The coordinator collects these outcomes and performs the verification test; if it fails, she increases the failure count $r_i$ by 1. Finally, she computes her local failure rate \(\delta_i = r_i/t_i\). More details are provided in End Matter~\ref{sec:Verification}.

\noindent\texttt{Voting.} When a Voting round is announced, each agent first checks that \(\delta_i \le \delta\); if not, they restart the subround from scratch. Every agent measures their qubit on the Hadamard basis. The agent whose secret index corresponds to the current round (the designated voter) XORs secret information (either randomness or a bit of their vote) into their measurement result. More details can be found in Protocol 2. Finally, all agents broadcast their possibly encoded measurement outcomes, forming a voting vector. At this point, any agent may invoke the \texttt{LogicalOr} subroutine with input `1' to abort if they detect an inconsistency between the broadcasted vector and their expectation. The collection of the $N$ voting vectors forms a sub-bulletin board, represented by a square in Fig.~\ref{fig:bulletin}.

To guarantee higher privacy, we compose \(\Pi\) repetitions of the protocol, a technique we refer to as privacy enhancement. These correspond to different columns in Fig.~\ref{fig:bulletin}. In each of the first \(\Pi - 1\) rounds, every voting agent casts a random vote $r_a$ and keeps track of it. In the final round, each voting agent casts a vote such that the parity of the XOR of all of their votes correctly encodes their voting preference.
Since the first \(\Pi - 1\) rounds contain no information about voting preferences, disclosing their measurement results does not leak any information. Privacy here refers to the probability that any malicious agent deviating from the honest protocol can guess any agent's vote is at most $\zeta$ more than in the case they just have access to the bulletin board and to their votes. The modifications introduced in the privacy enhancement and the overall structure of the protocol simplify its implementation and reduce the number of required samples. By allowing agents to locally abort subrounds—and making this decision available more often—we greatly improve efficiency compared to the original scheme. This makes the protocol compatible with high-fidelity quantum state sources, which typically operate at lower repetition rates~ \cite{martins2024realizingcompacthighfidelitytelecomwavelength}.

Similarly, to accommodate \(C\) candidates, we compose repetitions of several rounds (\( K = \lceil \log_2 C \rceil\)) of the protocol, where in each one the agents vote for the $k$-th bit of their voting preference expressed in binary units. This corresponds to different rows in Fig.~\ref{fig:bulletin}.
Finally, supporting multiple pools of voters does not require additional theoretical modifications: each pool can be treated as an independent protocol. The only practical assumption is that each pool may have its own security parameters and source characteristics. 
By running pools in parallel or sequentially, one can scale the system to larger electorates without altering the protocol. 


\begin{figure}[!htbp]
\centering
\begin{tikzpicture}[every node/.style={font=\small}]

  \node (matrix) at (0,0) {
    $\left(
    \begin{array}{c}
    \begin{tikzpicture}[every node/.style={font=\small}]
      \def\s{0.8}
      \def\gap{0.3}

      \foreach \i in {0,...,3} {
        \foreach \j in {0,...,4} {
          \coordinate (cell\i\j) at ({\j*(\s+\gap)}, {-\i*(\s+\gap)});
          \draw[thick] (cell\i\j) rectangle ++(\s,-\s);
        }
      }

      \node[above] at (0.5*\s, 0) {$p = 1$};
      \node[above] at ({4*(\s+\gap)+0.5*\s}, 0) {$p = \Pi$};
      \node[above] at ({2*(\s+\gap)+0.5*\s}, 0.2) {$\dots$};

      \node[left] at (0, -0.5*\s) {$k = 1$};
      \node[left] at (0, {-3*(\s+\gap)-\s+0.5*\s}) {$k = K$};
      \node[left] at (-0.2, {-1.5*(\s+\gap)-\s+0.5*\s}) {$\vdots$};

      \draw[fill=red,opacity=0.2] (-0.1,0.1) rectangle ++({5*(\s+\gap)-0.1},-1);
      \draw[fill=green,opacity=0.2] (-0.05,0.1) rectangle ++(0.9,{-4*(\s+\gap)+0.1});
    \end{tikzpicture}
    \end{array}
    \right)$
  };

  \node (matrices) at ($(matrix.south)+(0,-1)$) {
    $
    B_{k_{1},p_1} =
    \begin{bmatrix}
    1 & 1 & 1 & \textbf{0} \\
    \textbf{1} & 0 & 1 & 1 \\
    1 & 1 & \textbf{1} & 1 \\
    0 & \textbf{1} & 0 & 1
    \end{bmatrix}
    \quad \rightarrow \quad
    E_{k_{1},p_1} =
    \begin{pmatrix}
    1 \\
    1 \\
    0 \\
    0
    \end{pmatrix}
    \quad \rightarrow \quad
    T_{k_{1},p_1} =
    \begin{pmatrix}
    2 \\
    2
    \end{pmatrix}
    $
  };

  \draw[->, thick, red]
    ($(matrix.north west)+(1.6,-1.3)$)
    to[out=-200, in=150] ($(matrices.north)+(-3.3,-0.5)$);

\end{tikzpicture}
\caption{The bulletin of a the full protocol, with $\Pi$ privacy enhancement rounds, $K=\lceil \log_2 (C)\rceil$ bits to encode the $C$ possible candidates, and $N=4$ agents.
The red rectangle represents the first configuration, with two candidates and privacy enhancement; the green rectangle stands for the multi-candidate case with no privacy enhancement.
Each of the squares represents the sub-bulletin $B_{p,k}$, \textit{i.e.}, the $4 \times 4$ matrix generated for a fixed $p \in [\Pi]$ and $k \in [K]$. The election vector can be directly computed by evaluating the XOR of each line, and the tally simply counts the frequencies of the votes. The sub-bulletin $B_{p_1,k_1}$ represents a sub-bulletin of the election in voting order (3,0,2,1) for the first bit and first privacy enhancement round. The agents express their votes by adding them (0 or 1) to the row corresponding to their secret index (in bold), then broadcasting the resulting vector to form the sub-bulletin.
}
\label{fig:bulletin}
\end{figure}

\textbf{Experimental results.}\label{sec:Experimental result} To implement the protocol, we use a compact, high-fidelity four-party polarization-encoded GHZ state source based on spontaneous parametric down-conversion (SPDC) in a layered Sagnac interferometer configuration~\cite{martins2024realizingcompacthighfidelitytelecomwavelength}. The experimental setup is shown in detail in End Matter~\ref{sec:Setup}.
\begin{figure*}[!htbp]
\centering
{\includegraphics[width = \textwidth]{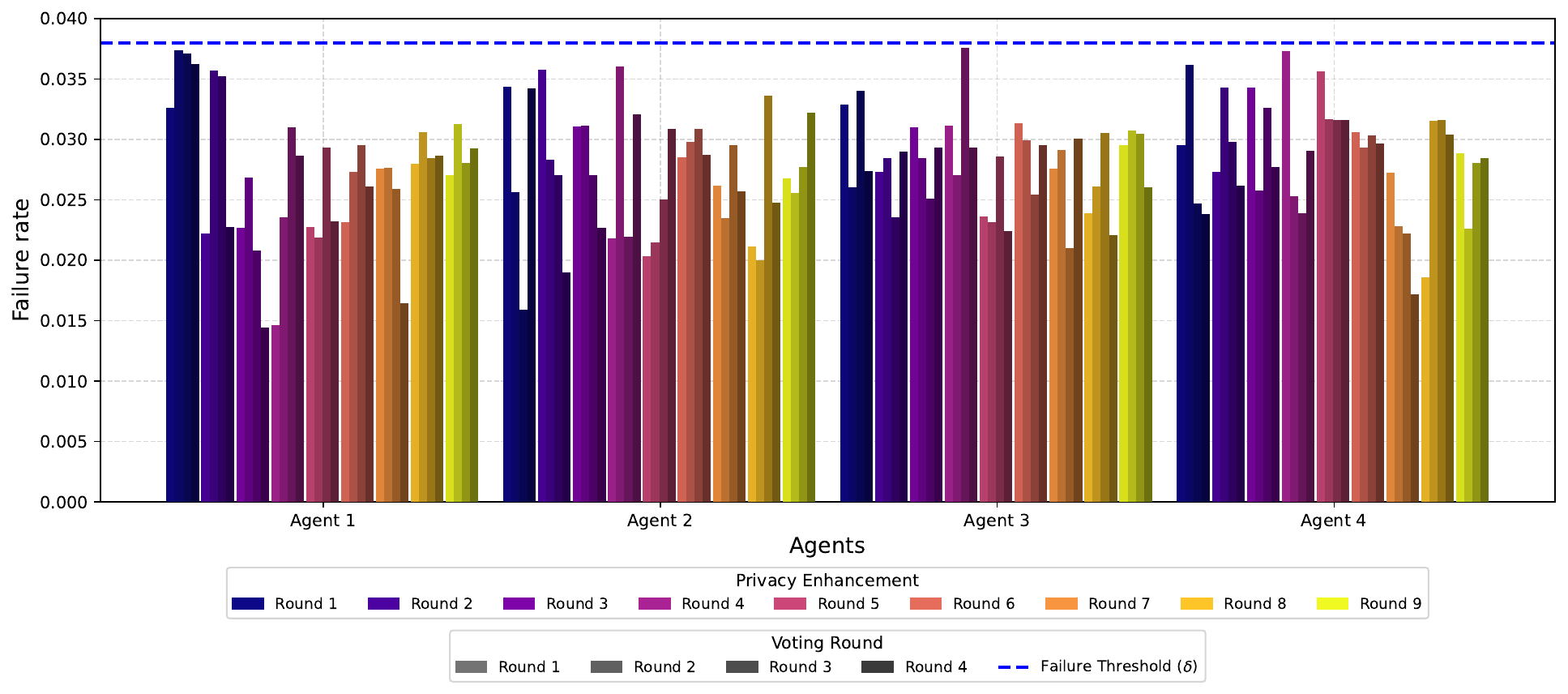}
\includegraphics[width = \textwidth]{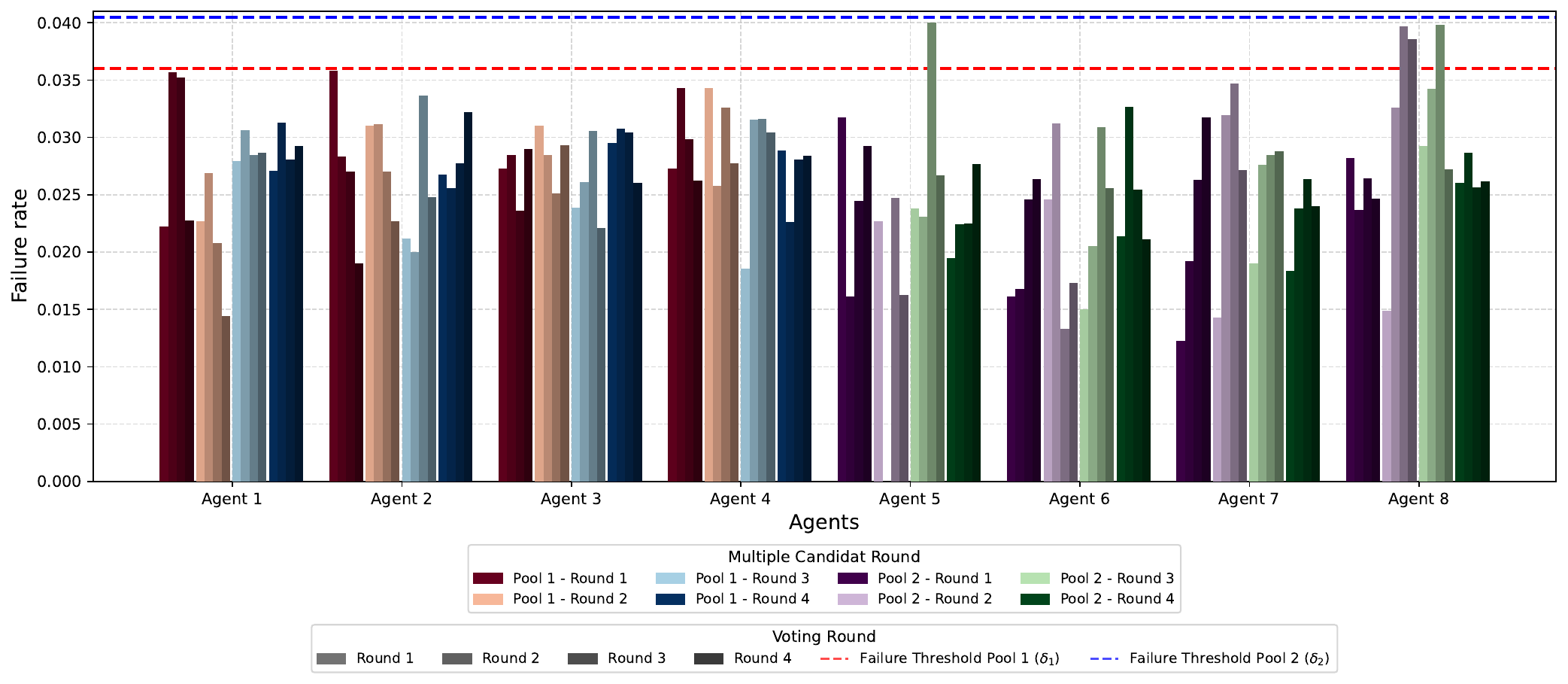}}
\vspace{-6mm}
\caption{Failure probability for the two scenarios under study. Each colour represents a different subroutine round while the shading for each colour indicates the individual voting rounds. The horizontal dashed line represents the failure rate threshold. \textbf{Top}: Scenario 1 with privacy enhancement. \textbf{Bottom}: Scenario 2 with multiple candidate rounds and voting pools.}
\label{fig:Scenario_1}
\end{figure*}

We begin our experimental analysis by performing quantum state tomography to estimate the intrinsic failure rate of our source. We generate a four‑qubit GHZ state with $96.27 \pm 0.46$ \% fidelity for an overall GHZ and Bell pair state generation per pulse of, respectively, $1,31\times10^{-7}$ and $2,59\times10^{-4}$. This high fidelity enables the protocol to be executed with high efficiency, thanks to a significantly low failure rate. The GHZ event rate of the source is roughly 10\,Hz. However, we cannot run the full protocol at this rate in practice. In the verification stage, each trial requires applying a randomly chosen rotation angle~$\theta$. We implement this unitary via a quarter-half-quarter wave‑plate sequence, but the mechanical rotation speed of each plate is slower than the source rate. Consequently, the end‑to‑end protocol rate is reduced to 1\,Hz. Although we could have reduced the GHZ event rate even further to increase the fidelity of our source, we chose to operate at a 10 Hz event rate. We found that this operating point provides the optimal experimental compromise between maintaining high state fidelity and achieving a practical protocol execution rate.

In the first protocol implementation scenario, we consider four agents voting between two candidates in a completely private manner. Through numerical analysis (see End Matter~\ref{sec:Simulation} for details), we determine that, for our source threshold of \(\delta = 0.0376\), a voting probability of \(2^{-M}\) with \(M = 13\) is required to 
guarantee a confidence level of 0.99. This in turn necessitates \(\Pi = 9\) rounds of privacy enhancement to expect privacy $\zeta = 0.01$. With these parameters fixed, we execute the election protocol. First, we confirm that in every verification round—and each voting subround—no agent’s observed failure probability \(\delta_j\) exceeds 0.0376 (Fig.~\ref{fig:Scenario_1}). This ensures that all participants remain within tolerance and none aborts due to excessive failure rates. Achieving this requires a sufficiently large \(M\) such that, with high probability, enough samples are gathered to avoid statistical anomalies. The fluctuations in \(\delta_j\) across rounds reflect both the varying sample sizes and the periodic unitary compensations we applied to counteract temperature drifts and mechanical shifts in the polarization controllers. Having passed all verification checks, we proceed to \texttt{Voting}. During each of the \(\Pi -1 \) privacy rounds, every voter casts a random vote $r_a$; only in the final round do they encode their final bit so that the XOR of all \(\Pi\) votes equals their intended bit.  The total bulletin \textbf{B} is provided in Supplemental Materials~\cite{supp}.
It is important to note that the voting vectors for each privacy enhancement round are revealed at the end of the corresponding round. This design makes the protocol more robust and experimentally manageable, as it allows for the immediate discarding of rounds where an error or an exceeded failure rate is detected.

The second protocol implementation scenario would ideally employ a second independent source to run the protocol in parallel; however, here we use the same source twice with deliberately different parameters. We omit privacy enhancement in this scenario, since the required sample size would be impractical. In Pool 1 (4 agents), we set \(M = 13\) and \(\delta = 0.036\), yielding 
and a confidence level of 0.99. In Pool 2 (also 4 agents), we use \(M = 12\), \(\delta = 0.0405\) 
maintaining the same 0.99 confidence level.

We perform the same verification analysis as in the first scenario. As shown in Fig.~\ref{fig:Scenario_1}, no verification round in either pool exceeds its respective failure threshold.
We publish the election bulletins and the final tally, demonstrating a true multi-candidate election with 8 voters. In our experiment, we simulate a contest among 16 candidates—comparable to recent French presidential primaries.
The total bulletin \textbf{B} is provided in Supplemental Materials~\cite{supp}. 

The practicality of this demonstration could be improved by adding privacy enhancement, which was not included here because of rates that would lead to an unrealistic duration for a voting procedure. This can be addressed in future work by improving the source rate or the wave-plate motor speed. A more fundamental practicality limitation arises from the small number of voters in each pool: a dishonest party can infer the collective result from the published bulletin alone. This vulnerability is intrinsic to decentralized voting schemes without a trusted authority, especially when the same entangled state is distributed to all voters. Nevertheless, recent advances in multi‑qubit GHZ‐state generation, demonstrating systems with more than 10 qubits~\cite{Thomas2022-uq,PhysRevLett.117.210502,PhysRevLett.121.250505}, suggest that scaling to larger voter pools may soon be feasible (see Supplementary Material~\cite{supp}).

Finally, we recall that a full implementation of the protocol would require quantum memories as agents can then store briefly their qubits until the chosen agent announces the measurement angle for verification or the voting trigger. In our experiments, we instead discard all photon counts that occur during the random selection between \texttt{Voting} and \texttt{Verification} and during the angular rotation of the wave‑plate assembly. However, quantum memories have recently been used for applications in quantum cryptography~\cite{ScienceAdvances2025}. Their incorporation into our setup is therefore an exciting and realistic future direction.

\textbf{Discussion.}\label{sec:Conclusion}
In this work, we have presented the experimental demonstration of our quantum electronic voting protocol within a rigorous theoretical framework allowing for privacy enhancement, supporting multiple candidates and voter pools, under realistic conditions that are aligned with currently available quantum technologies. We realized two distinct experimental scenarios. First, four agents cast their votes between two candidates without requiring any central authority, while maintaining vote privacy. This was only achievable by using our high-fidelity GHZ source capable of generating states with up to 96.27\% fidelity. Our results show that such a protocol is feasible for secure e-voting in near-term real-world applications. In the second scenario, we demonstrated an implementation involving multiple voting pools and multiple candidates, which we believe better reflects the structure of real-world elections. Modern elections are often organized by administrative pools (e.g., municipalities), making this configuration highly relevant. In our setup, we considered two pools of four voters each, voting among up to 16 candidates. This work marks a significant step forward toward practical quantum cities~\cite{2211.01190} by experimentally addressing a major protocol in quantum cryptography. To our knowledge, this constitutes the first faithful implementation of a quantum e-voting protocol. More generally, the ability to experimentally demonstrate quantum voting with such a high level of privacy paves the way for the reliable and robust deployment of quantum information systems in real-world settings.\\

\noindent\textbf{Note.} We acknowledge that a similar experiment was carried out independently by Marcellino \textit{et al.}~\cite{2512.03659}. While their implementation modifies the protocol to address source-trust assumptions without quantum memory, our work focuses on a new protocol to improve efficiency and scalability, while maintaining and demonstrating privacy. The complementarity between these approaches highlights the adaptability of our quantum electronic voting protocol and establishes several distinct paths toward real-world implementation.

\noindent\textbf{Data availability.} The developed code and data used for this work are available from the authors upon any reasonable request.

\noindent\textbf{Acknowledgments.} We thank Laura dos Santos Martins and Dominik Leichtle for fruitful discussions and technical support. We acknowledge financial support from the Horizon Europe research and innovation programme under the project QSNP (Grant No. 101114043) and the PEPR integrated project QCommTestbed (ANR-22-PETQ-0011), which is part of Plan France 2030.
\bibliography{references}
\clearpage


\section{Theoretical details}\label{sec:Simulation}
We study the main details of the protocol. Let us first formalize the security assumptions. The \texttt{Verification} subroutine is specifically designed to be robust against dishonest parties attempting to compromise the source by distributing malicious states \cite{PhysRevLett.108.260502}. Furthermore, the overall protocol is secure by design against a malicious or compromised central authority; the voters retain the ability to abort the protocol at any time if they detect discrepancies or anomalous behavior on the public bulletin, consequently they do not require any central authority. Finally, we address the requirement of a quantum memory, which would be essential to eliminate all trust assumptions regarding the state source. Without a quantum memory, a malicious party or compromised device could wait for the verifier to announce the required measurement bases and subsequently generate a forged state tailored to pass the verification for those specific angles. In such a scenario, one would be forced to blindly trust that the source is producing the real GHZ state. By requiring the voters to store their qubits in a quantum memory before the verifier's random angles are communicated, we can enforce a delayed-measurement regime. This would guarantee that the source cannot anticipate the measurement settings, thereby ensuring security without requiring a trusted source.

In the following we will focus, on \texttt{Verification}, to select the optimal parameters for our experimental implementation. The core equation governing the protocol is:
\begin{equation}
P(C_\epsilon) \leq \exp\left(-\frac{2^M(\epsilon^2 - 4\delta^2)}{16N\epsilon^2}\right)
\end{equation}
Here, $\delta$ represents the failure threshold: if the ratio of rejected rounds exceeds this value, the protocol is aborted. $P(C_\epsilon)$ is the probability that the protocol proceeds without aborting despite using a state that would fail the \texttt{Verification}, $M$ denotes the number of coin tosses an agent uses to decide between \texttt{Verification} and \texttt{Voting}, $N$ is the number of agents. Finally $\epsilon$ is the target infidelity we wish to verify. Our goal is to make $1 - P(C_\epsilon)$ as close to 1 as possible; we refer to this as the confidence level. 
We define the number of samples required per round as:
\begin{equation}
    2^M \geq \frac{16N\epsilon^2}{4\delta^2-\epsilon^2}\ln(P(C_\epsilon))
\end{equation}
This value must then be scaled by the failure probability of each protocol round :
\begin{equation}
1 - P_{\text{err}} \geq (1 - \delta)^4
\end{equation}
where $P_{\text{err}}$ is the probability to having an error per round. The privacy $\zeta$ is given by:
\begin{equation}
\zeta = (1 - \eta)^N \epsilon \sqrt{1 + \epsilon^2} + \left(1 - (1 - \eta)^N\right)
\end{equation}
with $\eta \geq P(C_\epsilon)$, (as introduced in the original paper~\cite{Centrone_2022}).
Strictly speaking, the \texttt{Verification} does not impose a direct bound on the absolute fidelity of the GHZ state. However, as detailed in~\cite{McCutcheon2016}, a strict upper bound on the allowable failure rate is required to successfully certify genuine multipartite entanglement (GME) and give an overall bound over the verified infidelity $\epsilon$. Therefore, the failure threshold $\delta$ is fundamentally constrained by the verification subroutine. Our experimental failure threshold remains well above this critical bound. 

The operational parameters of the protocol are deeply interconnected. Within the constraints of the GME, the experimental value of the failure threshold $\delta$ is deduced either from the density matrix or through initial experimental calibration runs. Based on this, we first choose the sample size; together with a specific confidence level, these parameters dictate the target infidelity $\epsilon$ that we are able to verify. While a naive approach to minimize the sample size would be to verify a higher infidelity, the verified state infidelity is the primary metric determining the protocol privacy bounds. A higher verified state infidelity yields inherently lower privacy. To achieve the desired level of privacy when operating with a higher infidelity, the protocol must be repeated $\Pi = \log(\zeta)/\log(\zeta_{wanted})$ times. Crucially, each of these $\Pi$ repeated rounds must be completed without a failure, which heavily penalizes the practical use of a lower-quality source. Furthermore, $1/2^M$ represents the probability of a voter casting a vote. Consequently, each voter undergoes an average of $2^M/N$ verification rounds. We must acquire sufficient data per voter to ensure that statistical fluctuations remain negligible. Finally, the minimum rejection threshold $\delta$ must be conservatively overestimated to account for experimental drift. Over extended periods, the source may experience instability and slow drift mainly cause by the fluctuation of the pump laser or polarization drift which naturally causes the baseline failure rate to slowly increase.

\section{Experimental setup}
\label{sec:Setup}
\begin{figure*}[!htbp]
\centering
{\includegraphics[width = 0.8\textwidth]{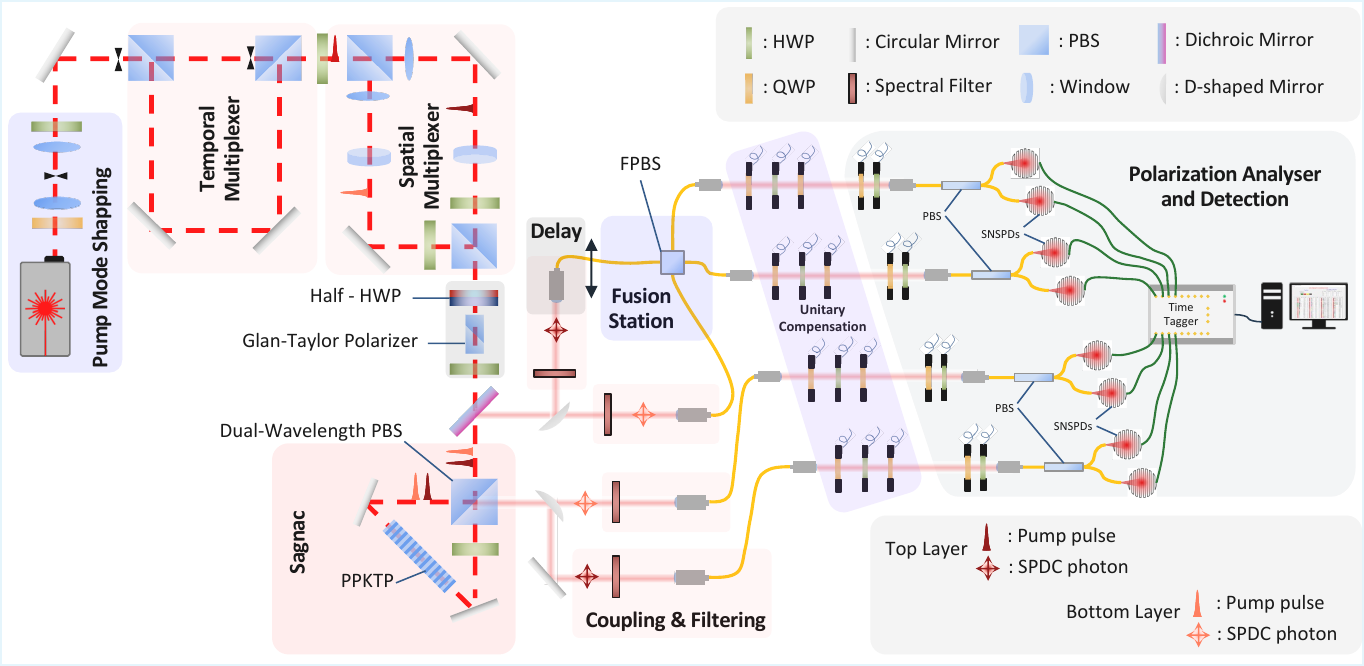}}
\caption{Layered Sagnac GHZ source. \textbf{Laser pump:} A Ti:Sapphire laser (Coherent Mira-HP) with an average power of 3.4~W emits 2~ps pulses at a wavelength of 775~nm with a repetition rate of 76~MHz. \textbf{Spatial mode shaping:} The spatial mode of the laser is shaped into a Gaussian profile. \textbf{Temporal multiplexer:} The pump pulses are split into two beams, the fast path and slow path. \textbf{Spatial multiplexer:} The pump pulses are split into two parallel beams: the top and bottom layers, horizontally and vertically polarized, respectively. \textbf{Polarization shaping:} Both layers are diagonally polarized to maximize the fidelity of output states concerning the Bell state. \textbf{Sagnac interferometer:} Photon pairs are probabilistically generated via type-II SPDC in a ppKTP crystal (30mm-long, 46.2~$\mu$m poling period, provided by Raicol) and entangled in polarization in the Sagnac loop, resulting in the output state $(\ket{H}_s\ket{V}_i+e^{i\theta}\ket{V}_s\ket{H}_i)/\sqrt{2}$. \textbf{Coupling and Filtering:} After filtering the single photons with a dichroic mirror, 1100~nm long pass filters and 1.3~nm ultra-narrowband filters, the bottom layer photons are reflected on half-circle shaped mirrors, while the top layer photons are transmitted over them. The photons are coupled to SM fibers with a 12~mm focal lens. \textbf{Fusion station:} The mechanical delay on the bottom-layer idler photon is fine-tuned such that both idler photons (from the top and bottom layers) arrive simultaneously at the FPBS. If each of them is transmitted to different outputs of the FPBS, and conditioned on fourfold coincidences, a GHZ state $(\ket{HHHH}+e^{i\delta}\ket{VVVV})/\sqrt{2}$ is generated. \textbf{Unitary compensation:} Three sets of QWP-HWP-QWP rotate the final state to the $\ket{GHZ}_{\delta=0}$ state. \textbf{Polarization Analyser and Detection:} A set of HWP-QWP-FPBS is used to map each photon's polarization to the spatial degree of freedom. The fibers are connected to superconducting nanowire single-photon detectors (ID281 SNSPD) that, in turn, are linked to a time tagger that allows for counting and correlating detection events for the analysis.}
\label{fig:Verifed_Fidelity}
\end{figure*}
\begin{figure*}
   \begin{tikzpicture}[scale=1, every node/.style={anchor=center}]
\centering
\coordinate (A) at (-1, 0);  
\coordinate (B) at (5, 0); 
\coordinate (C) at (10, 2);
\coordinate (D) at (10, -2);
\node at (A) {\scalebox{0.4}{\begin{tikzpicture}[every node/.style={align=center}]

    \def\rinner{1.2}  
    \def\router{4}    
    \def\rbig{2}      
    \def\n{6}         

    \fill[blue!5] (0,0) circle(\rbig); 
    \draw[blue!40!black, thick] (0,0) circle(\rbig);

    \foreach \i in {1,...,\n} {
        \node[circle, draw, fill=blue!30, minimum size=0.6cm] 
            (D\i) at ({\rinner*cos(360/\n*(\i-1))}, {\rinner*sin(360/\n*(\i-1))}) {};
    }

    \foreach \i in {1,...,\n} {
        \node[person, minimum size=1cm] 
            (P\i) at ({\router*cos(360/\n*(\i-1))}, {\router*sin(360/\n*(\i-1))}) {};

        \draw[->, thick] (D\i) -- (P\i);
    }

\end{tikzpicture}}};
\node at (B) {\scalebox{0.5}{\begin{tikzpicture}[every node/.style={align=center}, >={Latex[round]}]

    \def\n{6}
    \def\router{3}

    \foreach \i in {1,...,\n} {
        \node[person, minimum size=1cm] 
            (P\i) at ({\router*cos(360/\n*(\i-1))}, {\router*sin(360/\n*(\i-1))}) {};
    }

    \node[draw=red, thick, circle, minimum size=1.6cm, inner sep=0pt] at (P1) {};

    \node[below=3pt of P1] {\Large{Random Agent}};

    \node at ($(P1) + (1.8,0.4)$) (Verify) {\Large{Verify}};
    \node at ($(P1) + (1.8,-0.4)$) (Vote) {\Large{Vote}};

    \draw[->] (P1.east) -- (Verify.west);
    \draw[->] (P1.east) -- (Vote.west);

\end{tikzpicture}}};
\node at (C) {\scalebox{0.4}{\begin{tikzpicture}[every node/.style={align=center}, >={Latex[round]}]

    \def\n{6}
    \def\router{3}

    \foreach \i in {1,...,\n} {
        \node[person, minimum size=1cm] 
            (P\i) at ({\router*cos(360/\n*(\i-1))}, {\router*sin(360/\n*(\i-1))}) {};
    }
    
    \node[draw=blue, thick, circle, minimum size=1.6cm, inner sep=0pt] at (P4) {};
    \node[below=4pt of P4] {\Large{Verifier}};

    \foreach \i in {1,2,3,5,6} {
        \draw[<-] (P\i) -- node[near start, auto, font=\Large] {$\theta_i$} (P4);
    }

\end{tikzpicture}}};
\node at (D) {\scalebox{0.4}{\begin{tikzpicture}

    \def\n{6}
    \def\router{3}
    \def\moffset{1.2}

    \foreach \i in {1,...,\n} {
        \pgfmathsetmacro{\angle}{360/\n*(\i-1)}
        \pgfmathsetmacro{\x}{\router*cos(\angle)}
        \pgfmathsetmacro{\y}{\router*sin(\angle)}
        
        \node[person, minimum size=1cm] (P\i) at (\x,\y) {};
        
        \pgfmathsetmacro{\mx}{(\router+\moffset)*cos(\angle)}
        \pgfmathsetmacro{\my}{(\router+\moffset)*sin(\angle)}

        \ifnum\i=1
            \pgfmathsetmacro{\my}{\my + 0.3}
        \fi

        \node (M\i) at (\mx,\my) {
            \begin{quantikz}[transparent]
                \meter{}
            \end{quantikz}
        };

        \ifnum\i=1
            \node at (\mx, \my - 0.8) {\Large $\oplus$ $v$};
        \fi
    }

    \node[draw=red, thick, circle, minimum size=1.6cm, inner sep=0pt] at (P1) {};
    \node[below=3pt of P1] {\LARGE{Voter}};

\end{tikzpicture}}};
  \draw[->, thick] (1.2,0) -- (2.5,0);
  \draw[->, thick] (7.2,0.5) -- (8.3,1.3);
  \draw[->, thick] (7.2,-0.5) -- (8.3,-1.3);
\end{tikzpicture}
  
\end{figure*}
The polarization-entangled state used in our protocol is generated through the ``fusion'' of two Bell pairs. Each state is produced via type-II spontaneous parametric down conversion (SPDC) occurring within a periodically-poled KTP (ppKTP) crystal. To create the two Bell states, we split the pump into two parallel beams, using a spatial multiplexer, to pump the same ppKTP crystal in two different locations (top and bottom). Polarization entanglement is achieved by pumping the crystal from two opposing directions and then interfering the two paths using a Polarizing Beam Splitter (PBS), realized with a Sagnac interferometer~\cite{PhysRevA.73.012316}. As a result, we obtain the Bell state $\ket{\Phi} = (\ket{HV} + e^{i\theta}\ket{VH})/\sqrt{2}$, where $\theta$ is determined by the path difference between the two directions of propagation. We extract one photon from each pair and guide them to interfere on a Fiber Polarizing Beam Splitter (FPBS). Using a motorized delay stage, we finely adjust the temporal overlap of the interfering photons. We post-select the events resulting in each photon occupying a different spatial port of the FPBS. In other words, conditioned on fourfold coincidences, we entangle the two Bell pairs, thereby generating a GHZ state of the form $\ket{GHZ} = (\ket{HHHH} + e^{i\delta}\ket{VVVV})/\sqrt{2}$, where $\delta$ is determined by the $\theta$ of each Bell pair. With the aforementioned setup, we can generate a GHZ state up to local unitaries resulting from the propagation of the state in single-mode fibers. To specifically produce the state $\ket{GHZ}$, with $\delta=0$, we use an optimization method to determine the necessary local unitaries required to transform the state to the desired form. Using four sets of three Quarter-Wave Plates (QWPs), Half-Wave Plates (HWPs), and Quarter-Wave Plates (QWPs), we apply those unitaries to achieve the target state. More details can be found in~\cite{martins2024realizingcompacthighfidelitytelecomwavelength} and in the Supplementary Materials~\cite{supp}.

When \texttt{Verification} is performed, a verifier is anonymously selected to lead the verification procedure. During each voting round that includes a verification test (conducted before \texttt{Voting}), each honest agent $j$ keeps track of the number of verification trials and the number of rejections they observe. Based on this information, each agent independently computes their test statistic $\delta_j$. If $\delta_j$ exceeds the predetermined threshold $\delta$, the agent concludes that the verification has failed, and the round is aborted.
\section{Verification}
\label{sec:Verification}
\begin{protocol}{\texttt{Verification}}
\vspace{1pt}
\textit{Goal.} Verify if the parties share a GHZ state.
\sbline
\textit{Inputs.} The verifier label $v \in [N]$. Each voter has a qubit.
\sbline
\textit{Outputs.} Accept or reject.
\sbline
\textit{The protocol:}
\begin{enumerate}
  \item \textbf{Random angles.} The verifier generates and sends a random angle $\theta_j \in [0,\pi)$ to each voter $j \in [N]$, such that the sum $\sum_j \theta_j$ is a multiple of $\pi$.
  \item \textbf{Measurements.} Each agent $j$ measures their qubit in the basis $[\ket{+_{\theta_j}}, \ket{-_{\theta_j}}]$, and publicly announces the result $Y_j=\{0,1\}$.
  \item \textbf{Return.} If $\oplus_j Y_j= \frac{1}{\pi}\sum_j \theta_j$ return accept, otherwise return reject.
\end{enumerate}
\end{protocol}
 
\clearpage
\widetext
\begin{center}
\textbf{\large Supplemental Materials: Experimental Quantum Electronic Voting}
\end{center}

\begin{center}
\textbf{Classical subroutines}\\
\end{center}
For the sake of completeness, we provide the classical subroutines that are invoked in the End Matter. These are based on the original ideas of~\cite{Broadbent}, which were later developed in~\cite{Unnikrishnan_2019} and~\cite{Centrone_2022}.
A more detailed description of this protocol can be found in these papers.

The $\texttt{LogicalOR}$ subroutine is the most important one, from which all the others can be derived.
This allows a set of $N$ agents to privately compute the logical OR of their inputs, without requiring a simultaneous broadcast channel or a trusted central entity.

By tailoring the inputs of the agent, the $\texttt{LogicalOR}$ can be used to build $\texttt{RandomBit}$, where a fixed secret agent anonymously announces a random bit, and $\texttt{RandomAgent}$, where a fixed secret agent anonymously selects and announces a random agent.

Finally, the $\texttt{UniqueIndex}$ routine makes use of $\texttt{LogicalOR}$ in a slightly more complicated way, to distribute a secret ordering between the agents.
\begin{protocol}{\texttt{LogicalOR}}
\vspace{1pt}
\textit{Goal.} Publicly announce the OR of the variables of all parties, without revealing their private variables.
\sbline
\textit{Inputs.} $N$ participants. For $i \in [N]$, boolean variables $x_i\in\{0,1\}$. 
\sbline
\textit{Outputs.} Each agent gets $y_i = \bigvee_{i=1}^N x_i$.
\sbline
\textit{Parameters.} $S$ the security parameter.
\sbline
\textit{The protocol:}
\begin{enumerate}
  \item Choose $N$ random orderings, such that each ordering has a different last participant.
\item For each ordering, repeat $S$ times this procedure:
\begin{enumerate}
    \item Each agent $i$ generates a private bit $p_i$ as follows
    \begin{enumerate}
        \item if $x_i=0$, $p_i=0$;
        \item if $x_i=1$, then $p_i \in\{0,1\} $ is chosen uniformely at random.
    \end{enumerate}
    \item The agents collaboratively compute the parity $\oplus_{i=1}^N p_i$ in the following way:
    \begin{enumerate}
        \item each agent $i$ samples $N$ random bits $\{r_i^j\}_{j=1}^N$ such that $ \oplus_{j=1}^N r_i^j = p_i$;
        \item each agent $i$ sends bit $r_i^j$ to agent $j$;
        \item each agent $j$ computes $z_j=\oplus_{i=1}^N r_i^j$;
        \item following the fixed ordering, each agent publicly announces their value $z_j$;
        \item at the end, the value $z=\oplus_{j=1}^N z_j$ is computed, which corresponds to the bit $y_i$.
    \end{enumerate}
\end{enumerate}
\item Return $y_i=0$ if the result of the $N \times S$ repetitions is never $1$, otherwise return $y_i=1$.
\end{enumerate}

\end{protocol}

\begin{protocol}{\texttt{UniqueIndex}}
\vspace{1pt}
\textit{Goal.} Distribute a secret ordering between $N$ agents, without an authority.
\sbline
\textit{Inputs.} $N$ participants. For $i \in [N]$, boolean variables $x_i\in\{0,1\}$. 
\sbline
\textit{Outputs.} Each agent $i$ has a secret unique index $\omega_i \in [N]$.
\sbline
\textit{Parameters.} $S$ security parameter.
\sbline
\textit{The protocol:}
At the beginning, set $\omega_i=0$ for all agents.\\
For $R=1$ to $N-1$ :
\begin{enumerate}
  \item Each agent $i$ generates $x_i$ as follows:
  \begin{enumerate}
      \item if $\omega_i\neq 0$, $x_i=0$;
      \item if $\omega_i = 0$, then $x_i = 0$ with probability $1-\frac{1}{N-R}$, otherwise $x_i=1$.
  \end{enumerate}
  \item Perform the $\texttt{LogicalOR}$ of $\{x_i\}_i$ with security parameter $S$, and obtain the result $y_i$ :
  \begin{enumerate}
      \item If $y_i=0$, repeat the protocol from point 1.
      \item If $y_i=1$, at least one agent $k$ had input $x_k=1$. Set $c_i=0$ for everyone. All agents $k$ with $x_k=1$ check if they were the only one, i.e, if there were no collisions, by tracking the parity of all other inputs $\omega_k$. If for any of the $S$ repetitions in every ordering $\omega_k \neq 0$, set $c_k=1$.
      Perform a $\texttt{LogicalOR}$ of $\{c_i\}$:
      \begin{enumerate}
          \item If the result is $1$ (there was a collision), start again from 1.
          \item If the result is $0$ (there were not collisions), assign $\omega_k =R$ and go to the next iteration of the loop.
      \end{enumerate}
  \end{enumerate}
  \item At the end one agent $l$ will still have $\omega_l=0$, set it to $\omega_l=R$.
\end{enumerate}

\end{protocol}

\begin{protocol}{\texttt{RandomBit}}
\vspace{1pt}
\textit{Goal.} One fixed agent anonymously announces a bit sampled from a probability distribution $D$.
\sbline
\textit{Inputs.} Probability distribution $D$ with sample space $\Omega=\{0,1\}$.
\sbline
\textit{Parameters.} $S$ security parameter.
\sbline
\textit{The protocol:}
\begin{enumerate}
  \item The fixed agent $j$ sets $x_j$ to be $0$ or $1$ according to the distribution $D$; all other agents set $x_i=0$.
  \item Perform the $\texttt{LogicalOR}$ of the variables $\{x_i\}_i$, with security parameter $S$.
\end{enumerate}

\end{protocol}

\begin{protocol}{\texttt{RandomAgent}}
\vspace{1pt}
\textit{Goal.} One fixed agent randomly selects an agent, and anonymously announces it.
\sbline
\textit{Inputs.} $N$ agents, probability distribution $D$ with sample space $\Omega=[N]$.
\sbline
\textit{Parameters.} $S$ security parameter.
\sbline
\textit{The protocol:}
\begin{enumerate}
  \item The fixed agent $j$ samples a random variable $n = n_1 n_2 \dots n_K$ from the distribution $D$, which can be expressed in binary form with $K= \lceil \log_2(N)\rceil $ bits.
  \item For $k=1$ to $K$:
  \begin{itemize}
      \item The fixed agent $j$ sets $x_j^k = n_k$; all other agents set $x_i^k=0$.
      \item Perform the $\texttt{LogicalOR}$ of the variables $\{x_i^k\}_i$, with security parameter $S$.
  \end{itemize}
\end{enumerate}

\end{protocol}

\vspace{40 pt}

\begin{center}
\textbf{Additional experimental results}\\
\end{center}

\begin{figure}[!htp]
\begin{center}
\begin{tikzpicture}[scale=0.75]

\newcommand{\drawmatrix}[4]{
  \node at (#2 + 1.2, 1.5) {$B_{k_1,p_{#4}}$};
  \node at (#2 + 1.2, 0) {
    $\begin{array}{|cccc|}
    \hline
    #1
    \\ \hline
    \end{array}$};
  \node at (#2 + 0.5, -3) {
    $\mathbf{E}_{k_1,p_{#4}} = \begin{pmatrix} #3 \end{pmatrix}$};
}

\drawmatrix{\textbf{1} & 0 & 0 & 1 \\
            1 & 0 & \textbf{0} & 0 \\
            0 & 0 & 0 & \textbf{0} \\
            1 & \textbf{1} & 1 & 0}{0}{0 \\ 1 \\ 0 \\ 1}{1}

\drawmatrix{\textbf{0} & 1 & 0 & 1 \\
            1 & 0 & \textbf{0} & 0 \\
            1 & 1 & 1 & \textbf{1} \\
            1 & \textbf{0} & 1 & 0}{2.8}{0 \\ 1 \\ 0 \\ 0}{2}

\drawmatrix{\textbf{1} & 0 & 1 & 1 \\
            1 & 0 & \textbf{0} & 1 \\
            0 & 1 & 1 & \textbf{1} \\
            1 & \textbf{0} & 0 & 0}{5.6}{1 \\ 0 \\ 1 \\ 1}{3}

\drawmatrix{\textbf{0} & 1 & 1 & 1 \\
            1 & 1 & \textbf{1} & 0 \\
            0 & 1 & 0 & \textbf{0} \\
            1 & \textbf{1} & 1 & 1}{8.4}{1 \\ 1 \\ 1\\ 0}{4}

\drawmatrix{\textbf{1} & 0 & 1 & 1 \\
            0 & 0 & \textbf{0} & 1 \\
            1 & 1 & 0 & \textbf{0} \\
            0 & \textbf{1} & 0 & 0}{11.2}{1 \\ 1 \\ 0\\ 1}{5}

\drawmatrix{\textbf{1} & 1 & 0 & 0 \\
            0 & 0 & \textbf{0} & 1 \\
            0 & 1 & 1 & \textbf{0} \\
            0 & \textbf{0} & 1 & 0}{14.0}{0 \\ 1 \\ 0 \\ 1}{6}

\drawmatrix{\textbf{1} & 0 & 0 & 1 \\
            1 & 0 & \textbf{1} & 1 \\
            1 & 1 & 1 & \textbf{0} \\
            1 & \textbf{1} & 1 & 1}{16.8}{0 \\ 1 \\ 1 \\ 0}{7}

\drawmatrix{\textbf{1} & 1 & 1 & 0 \\
            0 & 0 & \textbf{0} & 1 \\
            0 & 1 & 1 & \textbf{0} \\
            0 & \textbf{0} & 0 & 0}{19.6}{1 \\ 1 \\ 0 \\ 0}{8}

\drawmatrix{\textbf{1} & 1 & 0 & 1 \\
            0 & 0 & \textbf{0} & 0 \\
            0 & 0 & 1 & \textbf{0} \\
            1 & \textbf{0} & 1 & 0}{22.4}{1 \\ 0 \\ 1 \\ 0}{9}

\node at (11.2, -6) {
  $\mathbf{T} = \begin{pmatrix} 2 \\ 2 \end{pmatrix}$
};
\end{tikzpicture}
\end{center}
\caption{Voting result for the first scenario: 4 agents who vote in the order (0, 2, 3, 1). The agnets express their vote by adding their vote (0 or 1) to the row corresponding to their secret index (in bold), then broadcast the resulting vector and all together they form the bulletin board $B_{k_n,p_{n}}$ . We have 8 privacy enhancement rounds where the agents voted randomly, and the last one where their encoded the vote depending on the precedent one. Here the votes were (1, 1, 0, 0). Then they sum each row of B to compute the election vote set $\mathbf{E}_{k_n,p_{n}}$, from which is computed the tally T.}
\label{tizx:Tally 1}
\end{figure}

\begin{figure}[!htp]
\centering
\begin{tikzpicture}[scale=1]

  \node at (1.5, 3.5) {\textbf{Pool 1}};
  \node at (9.5, 3.5) {\textbf{Pool 2}};

\node at (13.7, 3.5) {\textbf{Final Tally}};
  \draw[thick] (5, 4) -- (5, -5);
    \draw[thick] (12.3, 4) -- (12.3, -5);


  \newcommand{\drawblock}[4]{%
    \node[anchor=east] at (-1, #2) {$B_{k_{#4},p_{1}}=$};%
    \node at (0, #2) {%
      $\begin{array}{|cccc|}
        \hline
        #1 \\ \hline
      \end{array}$};
    \node[anchor=west] at (1.5, #2) {%
      $\mathbf{E}_{k_{#4},p_{1}}=
        \begin{pmatrix}
          #3
        \end{pmatrix}$};
  }

  \begin{scope}[xshift=0cm]
    \drawblock{0 & 0 & 1 & \textbf{0} \\ \textbf{0} & 1 & 1 & 1 \\ 1 & 1 & \textbf{1} & 1 \\ 1 & \textbf{0} & 0 & 0}{ 2}{1\\1\\0\\1}{1}
    \drawblock{1 & 1 & 0 & \textbf{1} \\ \textbf{1} & 1 & 1 & 1 \\ 0 & 1 & \textbf{1} & 1 \\ 1 & \textbf{1} & 0 & 0}{ 0}{1\\0\\1\\0}{2}
    \drawblock{0 & 0 & 1 & \textbf{1} \\ \textbf{0} & 0 & 1 & 1 \\ 0 & 1 & \textbf{0} & 1 \\ 1 & \textbf{0} & 0 & 1}{-2}{0\\0\\0\\0}{3}
    \drawblock{0 & 1 & 1 & \textbf{0} \\ \textbf{1} & 0 & 1 & 1 \\ 1 & 1 & \textbf{1} & 0 \\ 1 & \textbf{0} & 0 & 1}{-4}{0\\1\\1\\0}{4}
  \end{scope}

  \begin{scope}[xshift=8cm]
    \drawblock{0 & 1 & \textbf{1} & 0 \\ 0 & \textbf{1} & 1 & 1 \\ \textbf{1} & 1 & 1 & 0 \\0 & 0 & 0 & \textbf{1}}{ 2}{0\\1\\1\\1}{1}
    \drawblock{0 & 1 & \textbf{0} & 1\\ 0 & \textbf{0} & 0 & 0 \\ \textbf{1} & 0 & 1 & 1 \\ 1 & 0 & 1 & \textbf{0}}{ 0}{0\\0\\1\\0}{2}
    \drawblock{0 & 1 & \textbf{0} & 1 \\ 1 & \textbf{1} & 1 & 0 \\ \textbf{0} & 1 & 1 & 1 \\ 1 & 0 & 1 & \textbf{0}}{-2}{0\\1\\0\\0}{3}
    \drawblock{0 & 1 & \textbf{0} & 1 \\ 1 & \textbf{1} & 0 & 1 \\ \textbf{1} & 1 & 1 & 0 \\ 1 & 1 & 0 & \textbf{1}}{-4}{0\\1\\1\\1}{4}
  \end{scope}
  
  \node[anchor=west] at (12.5, -1) {
    $\mathbf{T} = 
      \begin{pmatrix}
        1 \\ 1 \\  0 \\ 1 \\  0 \\ 0 \\  0 \\ 0 \\ 0 \\ 2 \\  1 \\ 0 \\  
        1 \\ 0 \\  0 \\ 1 \\
      \end{pmatrix}$
  };
\end{tikzpicture}

\caption{Voting result for the second scenario: 4 agents for two pools who vote in the order (3, 0, 2, 1) and (2, 1, 0, 3). The agents express their votes by adding their votes (0 or 1) to the row corresponding to their secret index (in bold), then broadcasting the resulting vector. Together, they form the bulletin board $B_{k_n,p_{n}}$. We have 4 rounds where the agents vote to compose their 4 bits complete vote. Here, the votes were (3, 9, 10, 1) and (0, 13, 16, 9). Then they sum each row of B to compute the election vote set $\mathbf{E}_{k_n,p_{n}}$, from which the tally T is computed.}

\label{tizx:Tally 2}
\end{figure}

\vspace{40 pt}

\begin{center}
\textbf{Additional experimental details}\\
\end{center}

\begin{figure}[!htbp]
\centering
{\includegraphics[width = \textwidth]{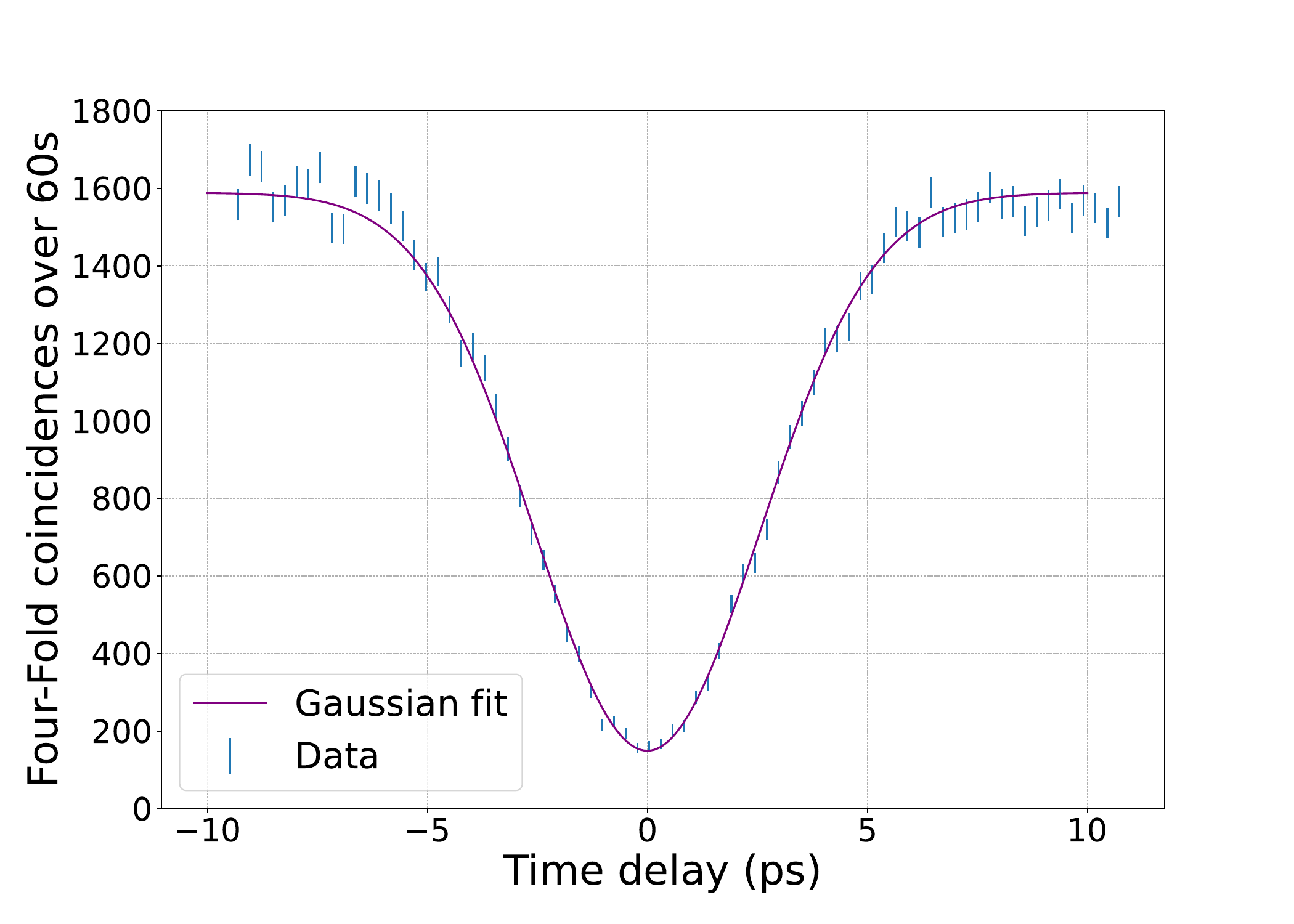}}
\caption{Hong-Ou-Mandel dip. The solid purple line is the Gaussian fit to
the experimental data (blue points), taken at full pump power, from which we
could infer a visibility of V = (90.62 ± 0.79)\%.}
\label{fig:HOM}
\end{figure}

\begin{figure}[!htbp]
\centering
{\includegraphics[width = \textwidth]{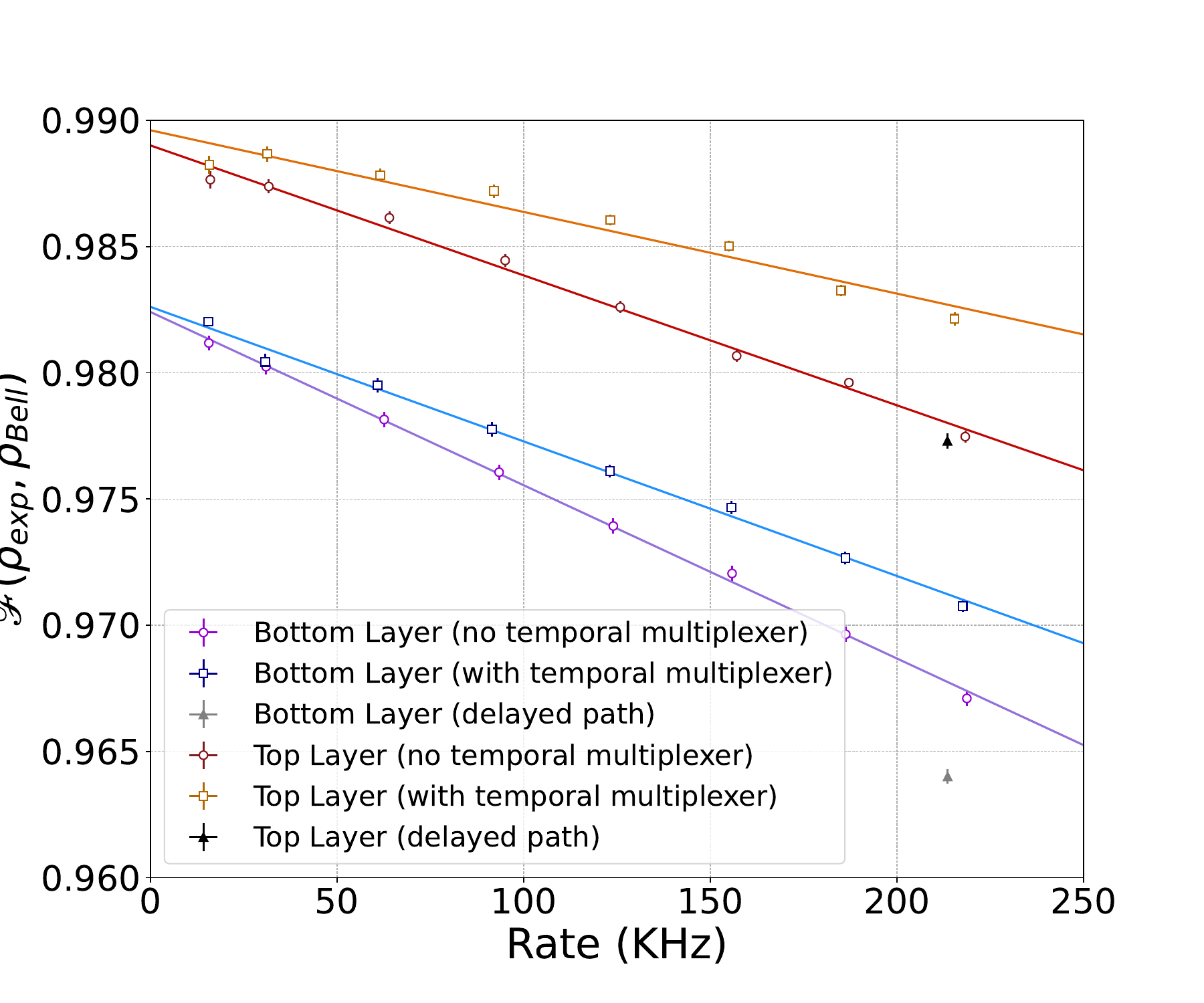}}
\caption{Experimental state fidelity with respect to the GHZ state, as a function of the two-fold coincidence rate. Results for the top layer, with and without the temporal multiplexer, are shown in orange and red, respectively. The same analysis for the bottom layer is depicted in blue (with the multiplexer) and purple (without the multiplexer). Linear fits to the experimental data are shown in the corresponding colours. The black and grey data points represent the cases where all pump light is delayed for the top and bottom layers, respectively. For the measurements involving the temporal multiplexer, 37\% and 87\% of the pump light is delayed for the top and bottom layers, respectively. The lower-rate blue and orange data points (leftmost points) correspond to pump powers of 25.6 mW and 35.8 mW, respectively.}
\label{fig:Temporal}
\end{figure}

To provide deeper experimental insight into our 4-qubit GHZ source, we include additional characterization data. First, we present the Hong-Ou-Mandel (HOM) interference dip of our GHZ state (Fig.~\ref{fig:HOM}). It is important to note that this HOM measurement was taken without the temporal multiplexer. Bypassing the multiplexer avoids introducing additional, variable delays, which significantly simplifies the alignment process for this specific characterization. Furthermore, to obtain the best possible counting statistics, this HOM measurement was performed at the maximum available pump power. Because of the increased multi-pair emission probability at this high power regime, the raw visibility of the HOM dip is naturally lower than the fully optimized state fidelity reported elsewhere in the text. Finally, following this characterization, we also present data explicitly detailing the effect of the temporal multiplexer, demonstrating how it improves the overall performance and equivalent fidelity of the source (Fig.~\ref{fig:Temporal}).

\begin{center}
\textbf{Scalability}\\
\end{center}

We now discuss the overall scalability of our protocol and the necessary steps toward realizing a large-scale quantum election. Expanding the system to accommodate more voters within a single pool requires increasing the number of qubits in the GHZ state. Using current photonic technologies, this intra-pool scaling presents significant experimental challenges~\cite{PhysRevLett.117.210502}. Generating larger entangled states compounds system losses and detector inefficiencies, which drastically reduces the viable GHZ event rate~\cite{martins2024realizingcompacthighfidelitytelecomwavelength}. Furthermore, the resulting degradation in state fidelity inherently raises the required failure threshold $\delta$, necessitating larger sample sizes to successfully complete the protocol. Consequently, the most viable path to a large-scale implementation is a hybrid, modular architecture. By optimizing a single source to its practical `sweet spot'- maximizing the number of qubits per pool while maintaining the requisite generation rate and fidelity bounds - we can scale the system horizontally. Multiplying these independent, optimized voting pools allows us to accommodate an arbitrarily large total number of voters, effectively bypassing the penalty of compounding quantum channel losses. Furthermore, recent demonstrations of high-visibility four-fold coincidences over long-range fibers indicate that long-distance multipartite quantum communication and GHZ-state generation are becoming practically viable~\cite{Zhan2025}.

\end{document}